\documentclass[preprint,12pt]{elsarticle}

\usepackage{amssymb}
\usepackage{amsmath}

\usepackage{tikz,pgfplots}
\usepgfplotslibrary{groupplots}
\usepackage{caption}
\usepackage{subcaption}
\pgfplotsset{compat=newest}
\usepackage{mathtools}
\usepackage{mathrsfs}
\usepackage{xcolor}
\usepackage{tikz,pgfplots}
\usetikzlibrary{shapes,arrows}
\usetikzlibrary{positioning}
\pgfplotsset{compat=1.18}
\usetikzlibrary{plotmarks}
\usetikzlibrary{arrows.meta}
\usepgfplotslibrary{patchplots}

\usepackage[a4paper, top=1in, bottom=1in, left=0.8in, right=0.8in]{geometry}

\usepackage{setspace}

\journal{arXiv}

\begin{document}

\begin{frontmatter}



\title{The Finite Element Neural Network Method: One Dimensional Study}


\author[poly]{Mohammed Abda}
\author[maya]{Elsa Piollet }
\author[maya]{Christopher Blake }
\author[poly]{Fr\'ed\'erick P. Gosselin}
\affiliation[poly]{organization={Laboratory for Multiscale Mechanics (LM2), Polytechnique Montréal},
            addressline={2500 Chemin de Polytechnique}, 
            city={Montréal},
            postcode={H3T 1J4}, 
            state={QC},
            country={Canada}}
\affiliation[maya]{organization={Maya HTT},
            addressline={1100 Av. Atwater Suite 3000},
            city={Montréal},
            postcode={H3Z 2Y4},
            state={QC},
            country={Canada}}

\begin{abstract}
The potential of neural networks (NN) in engineering is rooted in their capacity to understand intricate patterns and complex systems, leveraging their universal nonlinear approximation capabilities and high expressivity.  Meanwhile, conventional numerical methods, backed by years of meticulous refinement, continue to be the standard for accuracy and dependability. Bridging these paradigms, this research introduces the finite element neural network method (FENNM) within the framework of the Petrov-Galerkin method using convolution operations to approximate the weighted residual of the differential equations. The NN generates the global trial solution, while the test functions belong to the Lagrange test function space. FENNM introduces several key advantages. Notably, the weak-form of the differential equations introduces flux terms that contribute information to the loss function compared to VPINN, \textit{hp-}VPINN, and \textit{cv-}PINN. This enables the integration of forcing terms and natural boundary conditions into the loss function similar to conventional finite element method (FEM) solvers, facilitating its optimization, and extending its applicability to more complex problems, which will ease industrial adoption. This study will elaborate on the derivation of FENNM, highlighting its similarities with FEM.  Additionally, it will provide insights into optimal utilization strategies and user guidelines to ensure cost-efficiency.  Finally, the study illustrates the robustness and accuracy of FENNM by presenting multiple numerical case studies and applying adaptive mesh refinement techniques.
\end{abstract}



\begin{keyword}
Machine Learning \sep Neural network \sep Weak Formulation \sep Convolution \sep FEM\sep PINN \sep VPINN \sep \textit{hp-}VPINN \sep \textit{cv-}PINN \sep FENNM



\end{keyword}

\end{frontmatter}


\section{Introduction}\label{Introduction}
The Finite Element Method (FEM) remains the cornerstone of numerical simulation in engineering and applied mathematics due to its accuracy and reliability \cite{burnett1987finite, reddy1993introduction}. However, FEM requires well-posed problems with predefined parameters, loading, and boundary conditions. Moreover, its results are typically compared to experimental measurements only after the simulation. In contrast, a Physics-Informed Neural Networks (PINN) \cite{raissi2019physics,toscano_pinns_2024} seeks a solution by training a neural network on the appropriate physical laws, with the possibility to also use measurement data for the training. This versatility makes PINNs well-suited for solving ill-posed problems with incomplete, sparse, or noisy data while ensuring consistency with the underlying physics. Given the widespread success of FEM in handling complex physics and geometries, we aim to combine its strengths with the data-driven and inverse problem-solving capabilities of PINNs.

Neural netowrks (NNs) have been recognized since the 1990s \cite{dissanayake1994neural} as universal function approximators for solving differential equations. This is because of their ability to approximate a wide range of nonlinear functions using a relatively small number of parameters \cite{dissanayake1994neural, daubechies2022nonlinear, mhaskar2019function, jagtap2020conservative, kharazmi2021hp}. The numerical problem is formulated as an optimization problem in which the solution is approximated iteratively by minimizing a properly defined loss function \cite{Kharazmi2019}.  

This led to the development of PINNS, wherein the loss function is formulated from the strong-form residual of the differential equation evaluated at randomly distributed collocation points, constraining the network output toward the solution without requiring any high-fidelity data \cite{Raissi2017, Raissi2018_5, kharazmi2021hp, Kharazmi2019}.  PINNs offer a significant benefit over conventional solvers as they provide the solution as a function that is applicable throughout the entire domain, eliminating the need to create costly computational grids \cite{mcclenny2023self}. However, these vanilla PINNs have convergence and accuracy problems when solving stiff differential equations, solutions with sharp space transitions, and fast time evolution \cite{wang2020understanding}.

The variational formulation lowers the order of the differential equation, reducing the regularity required by the trial solution space, and making it better suited to handling stiff problems, singularities, and sharp changes \cite{Kharazmi2019, kharazmi2021hp}. Incorporating the variational formulation of the differential equation led to the development of variational physics-informed neural networks (VPINN) \cite{Kharazmi2019}, where random collocation points are replaced with Gauss quadrature points to approximate the weighted residual. The NN represents the nonlinear trial space of solutions, and the test functions are a combination of Legendre polynomials with vanishing values at the boundaries. Further development introduced \textit{hp-}refinement to VPINNs pushing it closer to FEM formulation by dividing the domain into nonoverlaping elements and using test functions with different orders \cite{kharazmi2021hp}. However, \textit{hp-}VPINN becomes expensive with increasing the number of elements, rendering it inapplicable to complex domains \cite{liu2023cv, anandh2024fastvpinns}. The computational cost is addressed by employing convolution operations in TensorFlow that parallelize the training process rather than looping over the elements in \textit{cv-}PINN \cite{liu2023cv}. However, the selection of test functions and the number of quadrature points required to approximate the variational formulation were not optimized, causing a loss in flux information between elements, adding unnecessary computational burden, and limiting their practical applications compared to FEM.

Recent work on variational PINN variants have demonstrated interesting properties and parallels with FEM, but without going all the way to merging PINN and FEM. Here we propose to do so and solve an FEM formulation with a neural network. We thus call this the \emph{finite element neural network method} (FENNM), which leverages the efficiency and precision of FEM and combines it with the flexibility and adaptability of variational PINNs based on the Petrov-Galerkin framework, taking advantage of the parallel capabilities of convolution operations in TensorFlow. The NN produces the nonlinear space of solutions, whereas the test functions belong to the Lagrange test function space with at least a nonvanishing value at the elements' boundaries. The weak form of the differential equation introduces flux terms along the interfaces, enabling FENNM to embed the natural boundary conditions and intermediate forcing terms into the residual loss function. Hence, FENNM takes a significant step closer to the classical FEM, narrowing the gap between machine learning and traditional numerical methods, and driving it towards industrial implementation. 

Section \ref{Theo} outlines the theoretical background that led to the development of FENNM. Following this, Section \ref{FENNM} presents a detailed formulation of FENNM. Section \ref{discussion} presents comprehensive numerical experiments aimed at analyzing in depth the impact of each component within the residual loss function on the design of FENNM solvers and examining how the convergence rate is affected by the order of the test functions. In addition, the accuracy and robustness of FENNM are evaluated in various benchmark problems. In Section \ref{conclusion}, the conclusion emphasizes the principal insights and prospective developments of FENNM within high-dimensional realms.
\section{Theoretical Background and Previous Work}\label{Theo}
\subsection{Finite Element Method (FEM)}\label{FEM}
The FEM is a mathematical technique that allows one to obtain numerical approximations for differential equations that represent physical systems that are usually subjected to external loads \cite{burnett1987finite}. One of the strengths of FEM is its ability to provide approximations to complex problems that are difficult to solve using other methods. This is because the finite element solution can be used repeatedly for all elements in the same mesh and adapted to different problems with minimal modifications \cite{burnett1987finite}.  
Consider the following illustration problem shown in equation (\ref{FEM_eq}), also called the \textit{equilibrium problem}
\begin{equation}\label{FEM_eq}
\frac{d}{dx}\left( x\frac{dU(x)}{dx} \right) = \frac{2}{x^2},\ \ \  x = [1,2],
\end{equation}
with $x$ as the independent variable. The essential and natural boundary conditions are, respectively 
\begin{equation}\label{FEM_eq_BC}
U(1) = 2, \ \ \  \\
\tau (2) =  \left .\left(-x\frac{dU}{dx}\right)\right |_{x=2} = \frac{1}{2}.
\end{equation}

In FEM, the solution to a differential equation is approximated by choosing a trial function from a finite-dimensional space and reducing the residuals of the equation by weighting them with a set of test functions. The trial functions and the test functions belong to a linear space where they are the same in the Galerkin framework and are distinct in the Petrov-Galerkin framework, producing different numerical schemes depending on the choice of the trial functions \cite{burnett1987finite, Kharazmi2019, reddy1993introduction}. The FEM formulation of equation (\ref{FEM_eq}) using the Galerkin method for one element is \cite{burnett1987finite}
\begin{equation}
     \left. \phi_k(x)x\frac{d\hat{U}}{dx}\right|_{x_n}^{x_{n+1}} -  \int_{x_n}^{x_{n+1}} \frac{d \phi_k(x)}{dx}   x\frac{d\hat{U}(x)}{dx} dx - \int_{x_n}^{x_{n+1}} \phi_k(x) \frac{2}{x^2}  dx = 0,
\end{equation}
where $\phi_k (x)$ is the  $k_{th}$ test function, and $\hat{U}(x)$ is the trial solution. The flux term is evaluated at the element left boundary $x_n$ and the right boundary $x_{n+1}$. Equation (\ref{FEM_eq}) will be used as an example to demonstrate the concepts discussed in the current work.  

In a more general setting, nonlinear trial functions extend the approximations to a nonlinear space, resulting in a more robust estimation with sparser representation and reduced computational cost \cite{devore1998nonlinear, devore2009nonlinear}. Nonlinear approximation approaches can include radial basis functions \cite{devore2010approximation}, dictionary learning \cite{tariyal2016greedy}, neural networks \cite{daubechies2022nonlinear}, and adaptive splines \cite{devore1998nonlinear}. However, while nonlinear approximation introduces additional capabilities, its nonlinear nature brings additional complications, and achieving an optimal approximation rate can become challenging, especially in high-dimensional spaces \cite{Kharazmi2019}. 
\subsection{Physics Informed Neural Networks (PINN)}
NNs transform high-dimensional input into output through algebraic operations and nonlinear mapping \cite{daubechies2022nonlinear, mhaskar2019function}. They function as an optimization method by iteratively adjusting their parameters to minimize a loss function that measures the discrepancy between the network output and high-fidelity data. A key advantage of NNs is their ability to represent a wide range of nonlinear functions using a relatively small number of parameters \cite{Kharazmi2019}. 

Although NNs are not inherently data-driven, when used to fit data, they are unaware of the mathematical model expressing physical laws. Hence, they require a large amount of high-fidelity data to achieve accurate and reliable predictions. This becomes problematic in small-data regimes, where the available data is insufficient relative to the complexity of the system \cite{raissi2019physics,Kharazmi2019}.  Constructing physics-informed learning machines replaces the large data requirement by embedding the prior information of the differential equations into NNs , which  led to the development of PINNs \cite{Raissi2017, Raissi2018_5}.

PINN includes the differential equation as a residual term at random collocation points in the computational domain, which acts as a penalizing term constraining the space of solutions \cite{Raissi2017, Raissi2018_5, kharazmi2021hp, Kharazmi2019}. Hence, inferring the solution of the differential equation is transformed into an optimization problem of the residual term which acts as a loss function at the penalizing points generated at minimal cost. To construct a PINN for the \textit{equilibrium problem} (\ref{FEM_eq}), the loss function is expressed in the strong-form for equation (\ref{FEM_eq}) and equation (\ref{FEM_eq_BC}) as
\begin{equation}\label{FEM_eq_PINN}
\begin{split}
\mathcal{L} &= \tau_\mathcal{R}\mathcal{L}_\mathcal{R}  + \tau_\mathcal{B}\mathcal{L}_\mathcal{B},  \\
\mathcal{L}_\mathcal{R} &= \frac{1}{N_{\mathcal{R}}} \sum_{i=1}^{N_{\mathcal{R}}} \left ( \frac{d}{dx}\left( x\frac{dU_{NN}(x)}{dx} \right) - \frac{2}{x^2}\right)^2, \\
\mathcal{L}_\mathcal{B} &= \frac{1}{N_{\mathcal{B}}} \sum_{i=1}^{N_{\mathcal{B}}}((U_{NN}(1) - 2)^2 + \frac{1}{N_{\mathcal{B}}} \sum_{i=1}^{N_{\mathcal{B}}} \left (\left .\left(-x\frac{dU_{NN}}{dx}\right)\right |_{x=2} - \frac{1}{2}\right)^2,
\end{split}
\end{equation}
where $\tau_{\mathcal{R}},\tau_{\mathcal{B}}$ are penalty parameters for the residual loss term $\mathcal{L}_\mathcal{R}$ and the boundary conditions loss term $\mathcal{L}_\mathcal{B}$, respectively. The penalty parameters can be manually or automatically adjusted during training \cite{mcclenny2020self}. The collocation points $N_{\mathcal{R}}$  are randomly sampled constructing the computational domain, and the number of boundary points $N_{\mathcal{B}}$ is chosen before training.

Despite the potential of PINNs in tackling forward problems, they suffer from limitations that hinder their efficiency compared to classical methods like FEM. The loss function of the network consists of different terms as shown in equation (\ref{FEM_eq_PINN}) that can cause convergence problems as optimization becomes highly non-convex \cite{blum1988training}. Furthermore, vanilla PINNs struggle to converge and provide an accurate approximation for stiff problems that contain solutions with sharp changes in space \cite{burden2010numerical}. 

Some techniques address the computational domain of PINNs, such as adaptive sampling strategies based on residual-based adaptive distribution \cite{wu2023comprehensive, mao2023physics}. Other methods divide the spatial domain into discrete subdomains such as conservative PINNs (cPINN), in which separate PINNs are applied in each subdomain while enforcing the flux continuity along the interfaces \cite{jagtap2020conservative}. The parallization powers of cPINN are extended in eXtended PINNs (XPINN) to the spatial and temporal domains for all types of differential equations, reducing training and computational costs \cite{jagtap2020extended}. However, in all these formulations, the strong form of mathematical models is employed at random collocation points, requiring a large number of points to guarantee convergence \cite{liu2023cv}. Although cPINN and XPINN provide parallization capabilities, they introduce additional layers of complexity. The use of multiple networks complicates the hyperparameter tuning for each network and the challenge of connecting the networks along the interfaces. 
  
\subsection{Variational Physics Informed Neural Networks (VPINN)}\label{VPINN}
Incorporating the weighted residual of the differential equation to construct a variational loss function results in VPINNs. The loss function is developed within the Petrov-Galerkin framework, where the test functions belong to a linear space and are a combination of Legendre polynomials, while the nonlinear approximation of the NN represents the trial solution \cite{Kharazmi2019}. The weak formulation reduces the regularity required in the network output by lowering the operator orders in the loss function. This approach reduces automatic differentiation computations when generating the loss terms and reduces the computational cost by minimizing \textit{backpropagation} processes.  \cite{Kharazmi2019,kharazmi2021hp}. The residual loss term of equation (\ref{FEM_eq_PINN}) becomes
\begin{equation}\label{VPINN_for_FEM_eq}
    \mathcal{L}_\mathcal{R} = \frac{1}{K} \sum_{k=1}^{K}  \left (\int_{\Omega}\left( \phi_k(x) , \frac{d}{dx}\left( x\frac{dU_{NN}(x)}{dx} \right) - \frac{2}{x^2}\right) \partial \Omega \right)^2,
\end{equation}
where $(,)$ indicates the inner product between the $k_{th}$ test function $\phi(x)_k,$$\ k = 1,2,\dots K$ and the residual of the differential equation. A version of VPINN is the variational neural network (VarNet), which employs piecewise linear test functions of FEM in the variational formulation \cite{khodayi2020varnet}. Both VPINN and VarNet compute the weak-form integral over the whole domain, which reduces their approximating capabilities in complex domains.

VPINNs were extended to consider the domain decomposition in  \textit{hp}-VPINN, where the test functions are defined locally over nonoverlapping elements, and the NN represents the global nonlinear trial solution \cite{kharazmi2021hp}. The \textit{hp}-refinement provides the flexibility of domain decomposition by applying \textit{h}-refinement using variable element sizes with a projection onto a space of high order polynomials as \textit{p}-refinement, making \textit{hp}-VPINN the first method to approximate solutions like FEM by discretizing the domain into elements and incorporating \textit{hp}-refinement techniques \cite{kharazmi2021hp}. Equation (\ref{VPINN_for_FEM_eq}) becomes
\begin{equation}\label{hp-VPINN_for_FEM_eq}
    \mathcal{L}_\mathcal{R} = \frac{1}{N_{el}K}\sum_{n=1}^{N_{el}} \sum_{k=1}^{K^{(n)}}  \left (\int_{\Omega}\left( \phi^{(n)}_k(x^{(n)}) , \frac{d}{dx}\left( x^{(n)}\frac{dU_{NN}(x^{(n)})}{dx} \right) - \frac{2}{x^{2(n)}}\right) \partial \Omega \right)^2,
\end{equation} 
where $N_{el}$ is the number of elements. There is no analytical solution for the integrals of the weighted residuals, and numerical integration techniques such as the Gauss quadrature rule is used. Hence, the residual loss in equation (\ref{hp-VPINN_for_FEM_eq}) is approximated to
\begin{equation}\label{q-hp-VPINN_for_FEM_eq}
    \mathcal{L}_\mathcal{R} = \frac{1}{N_{el}K}\sum_{n=1}^{N_{el}} \sum_{k=1}^{K^{(n)}}  \left (\sum_{q=1}^{Q}\left(W_q \phi_k (x_q^{(n)}) , \frac{d}{dx}\left( x_q^{(n)}\frac{dU_{NN}(x_q^{(n)})}{dx} \right) - \frac{2}{x_q^{2(n)}}\right)\right)^2,
\end{equation} 
where the $q_{th}$ quadrature point is located at position $x_q$ in the element and $W_q$ is the corresponding quadrature weight. The test functions can be defined in local or global coordinates and the necessary transformations must be applied when evaluating the residual loss function. There are fewer quadrature points than collocation points in a regular vanilla PINN, which decreases the computational cost \cite{berrone2022variational}. Increasing the number of elements is more efficient than adding more quadrature points within each element \cite{liu2023cv}. However, increasing the number of elements exponentially increases the computational cost, rendering \textit{hp}-VPINN inefficient for approximating complex functions and thereby limiting its practical applications \cite{liu2023cv, anandh2024fastvpinns}.
\subsection{Convolutional Variational Physics Informed Neural Network (\textit{cv}-PINN)}
The computational cost in \textit{hp}-VPINN is addressed by applying convolution operations to compute the strong-form weighted residual loss in \textit{cv}-PINN. The product of the test functions and the quadrature weights form convolution filters passing over the strong-form residuals \cite{liu2023cv}. By evaluating the loss function with convolution operations in TensorFlow, rather than sequentially looping over elements for each test function as done in \textit{hp}-VPINN, \textit{cv}-PINN gains the advantage of parallelizing the training process. The residual loss term can be written as
\begin{equation}\label{cVPINN_for_FEM_eq}
    \mathcal{L}_\mathcal{R} = \frac{1}{N_{el}K}\sum_{n=1}^{N_{el}} \left(\boldsymbol{c_{k}^{(n)}} , \frac{d}{dx}\left( x_q^{(n)}\frac{dU_{NN}(x_q^{(n)})}{dx} \right) - \frac{2}{x_q^{2(n)}}\right)^2,
\end{equation} 
where $\boldsymbol{c_k^{(n)}} = W_q \phi_k(x_q^{(n)})$ is a matrix of the convolution filters. As shown in Equation (\ref{cVPINN_for_FEM_eq}), the convolution operation simultaneously performs the integral approximation using the Gauss quadrature rule on all test functions.

However, VPINN, \textit{hp-}VPINN, and \textit{cv-}PINN have a significant limitation. They rely on the use of a set of Legendre polynomials as test functions, which vanish at the element boundaries, leading to a loss of flux information across the elements \cite{Kharazmi2019, kharazmi2021hp, liu2023cv}. Moreover, they employ a large number of high-order test functions and quadrature points, failing to optimize their selection and, thereby, adding additional computational burden. The order of the test functions and their influence in selecting an adequate number of quadrature points remained an open question.
\section{The Finite Element Neural Network Method (FENNM)}\label{FENNM}
\begin{figure}
\centering
\normalfont
\def\svgwidth{43cm}
\hspace*{-1.6cm}
\input{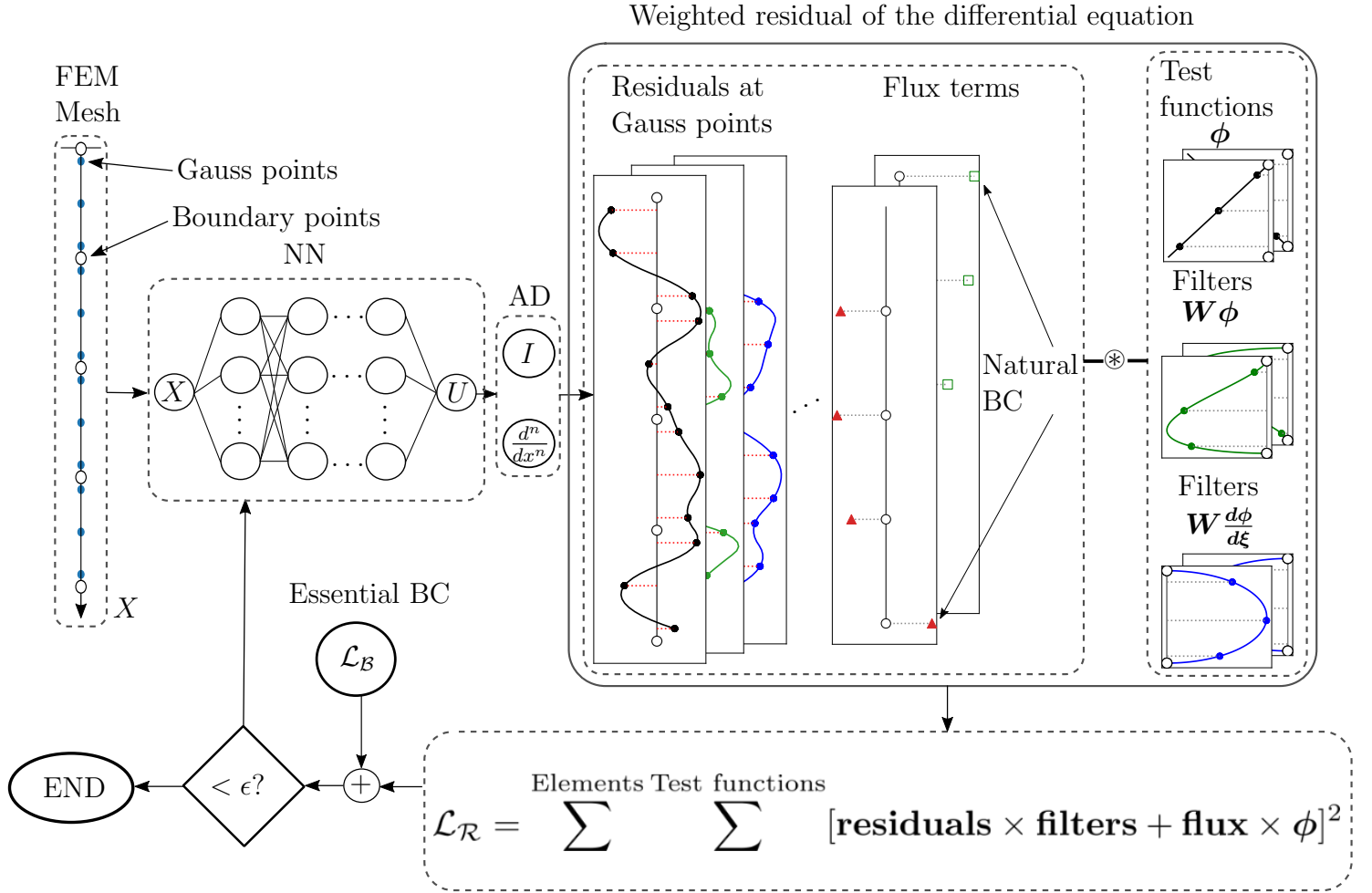}
\caption{Schematic of one-dimensional FENNM. The NN outputs formulate the fluxes, differential equation operators, and forcing terms. Then predefined filters pass over them to compute the Gauss quadrature sums in the convolution process before evaluating the global residual loss function. }
\label{FENNM_schematic}
\end{figure}

To overcome the challenges mentioned above, the current work presents the finite element neural network method FENNM using the convolution operations introduced in \textit{cv-}PINN. The NN provides the global nonlinear space of solutions, while the test functions belong to the Lagrange test function space, and have at least one nonvanishing value at the element boundaries. Consequently, the information of the flux terms across the elements is now implemented inside the weak-form loss function. Figure \ref{FENNM_schematic} presents a schematic of the one-dimensional FENNM solver. In the automatic differentiation step, the NN output generates the fluxes, weak-form residuals of the differential equation including the forcing term using automatic differentiation termed signals. The predefined filters comprising the test functions and their derivatives and quadrature weights then pass over these signals in the convolution process to construct the residual loss. Finally, the total residual loss is evaluated and the NN parameters are updated iteratively to minimize the total loss. 

 Consider equation (\ref{cVPINN_for_FEM_eq}); the weak-form per element for the $k_{th}$ test function after integrating by parts is
\begin{equation}\label{FENNM_formulation_1}
    \mathcal{L}_\mathcal{R}^{(n)} = \left. \phi_k(x)x\frac{dU_{NN}}{dx}\right|_{x_n}^{x_{n+1}} -  \int_{x_n}^{x_{n+1}} \frac{d \phi_k(x)}{dx}   x\frac{dU_{NN}(x)}{dx} dx - \int_{x_n}^{x_{n+1}} \phi_k(x) \frac{2}{x^2} dx. 
\end{equation}

As convolution filters are fixed and pass over the NN output, it is essential to define the test functions in local coordinates. This approach allows for the generalization of the formulation to accommodate meshes with adaptive element sizes similar to FEM.  Hence, equation (\ref{FENNM_formulation_1}) becomes
\begin{equation}\label{FENNM_formulation_2}
    \mathcal{L}_\mathcal{R}^{(n)} = \left. \phi_k(\xi)x\frac{dU_{NN}(x)}{dx}\right|_{x_n}^{x_{n+1}} -  \int_{-1}^{1} \frac{d \phi_k(\xi)}{d\xi}   x\frac{dU_{NN}(x)}{dx} d\xi - \int_{-1}^{1} \phi_k(\xi) \frac{2}{x^2}  J_x d\xi,
\end{equation}
where $\xi$ denotes the local coordinate within the interval $[-1,1]$ such that $x = x_n + J_x (1+\xi )$, and $J_x = (x_{n+1}-x_n)/2$ is the one-dimensional Jacobian. The integral terms in equation (\ref{FENNM_formulation_2}) are approximated using Gauss quadrature rule as
\begin{equation}\label{FENNM_formulation_3}
     \mathcal{L}_\mathcal{R}^{(n)} = \left. \phi_k(\xi)x\frac{dU_{NN}(x)}{dx}\right|_{x_n}^{x_{n+1}} -  \sum_{q=1}^Q W_q \frac{d \phi_k(\xi_q)}{d\xi}   x_q\frac{dU_{NN}(x_q)}{dx}  -   \sum_{q=1}^Q J_x  W_q \phi_k(\xi_q) \frac{2}{x_q^2}.
\end{equation}

An additional benefit for employing Lagrange test functions is that they allow for the integration of natural boundary conditions directly into the residual loss function. This integration smooths the total loss function, thus improving the convergence and optimization of FENNM. Moreover, the nonvanishing test function enables the application of intermediate forcing terms within one neural network. The residual loss function for $N_{el}$ elements and $K$ test functions becomes 
\begin{equation}\label{FENNM_residual_eq}
    \begin{split}
        \mathcal{L}_\mathcal{R} &= \frac{1}{N_{el} K} \sum^{N_{el}}_{n=1} \sum^K_{k=1} 
        \Bigg( \phi_k(\xi) \underbrace{\left. x \frac{dU_{NN}(x)}{dx} \right|_{x_n}^{x_{n+1}}}_{\text{Fluxes}}
        - \sum_{q=1}^Q \underbrace{W_q \frac{d \phi_k(\xi_q)}{d\xi}}_{\text{Filters1}} 
        \underbrace{x_q^{(n)} \frac{dU_{NN}(x_q^{(n)})}{dx}}_{\text{DE operators}} \\
        &\quad - \sum_{q=1}^Q J_x \underbrace{W_q \phi_k(\xi_q)}_{\text{Filters2}} 
        \underbrace{\frac{2}{x_q^{2(n)}}}_{\text{Forcing term}} \Bigg)^2.
    \end{split}
\end{equation}

The loss function in equation (\ref{FENNM_residual_eq}) consists of four parts:
\begin{itemize}
    \item Fluxes: arise at the boundaries on both the left and right sides of each element defined in the global coordinates.
    \item DE operators: the differential equation operators with lower orders after integration by parts evaluated in the global coordinates.
    \item Filters: which are the product of the Gauss quadrature weights and the test functions or their derivatives according to the number of integration by parts performed such as in Filters1 and Filters2 in equation (\ref{FENNM_residual_eq}). These filters are computed in the local coordinates to ensure their generalizability over elements of different sizes. The formulation of the filters is demonstrated in Figure \ref{filters} using linear Lagrange test functions as an example. Filters1 are represented in (a) and (b), while Filters2 are represented in (c) and (d).
    \newcommand{\blackbullet}{\raisebox{0pt}{\tikz{\filldraw[black] (0,0) circle (1pt)}}}
    \newcommand{\bluesquare}{\raisebox{0pt}{\tikz{\filldraw[blue] (0,0) rectangle (4pt,4pt)}}}
    \definecolor{darkgreen}{rgb}{0.0, 0.5, 0.0}
    \newcommand{\greenhollowsquare}{\raisebox{0pt}{\tikz{\filldraw[darkgreen] (0,0) rectangle (4pt,4pt)}}}
    \definecolor{darkgreen}{rgb}{0.0, 0.5, 0.0}
    \newcommand{\greenbullet}{\raisebox{0pt}{\tikz{\filldraw[darkgreen] (0,0) circle (2pt)}}}
    \begin{figure}
    \centering
    \includegraphics[width=1.\linewidth]{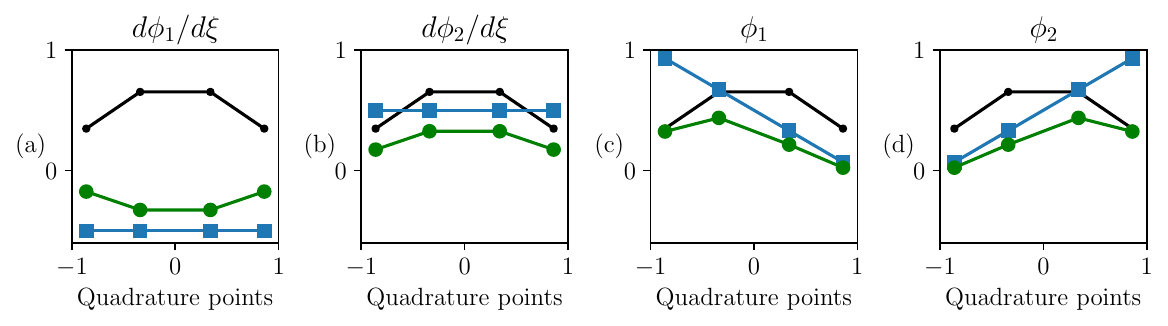}
    \caption{Construction of the convolution filters for linear Lagrange test functions with $Q = 4$ Gauss points: (a) Filter $W_q d\phi _1/d\xi$; (b) Filter $W_q d\phi _2/d\xi$; (c) Filter $W_q \phi _1$; (d) Filter $W_q \phi _2$. Markers indicate the positions of Gauss points: \protect\blackbullet  \ integration weights $W_q$; \protect\bluesquare  \ test functions and their derivatives; \protect\greenbullet 
 \ filters.} \label{filters}
\end{figure}
    \item Forcing term: includes any source terms and the remaining DE operators evaluated in the global coordinates. 
\end{itemize}

The global coordinates define the computational domain, establishing a reference frame for the elements, the quadrature points, the elements' boundaries, and the boundary conditions. Figure \ref{uniform_mesh} shows a one-dimensional grid divided uniformly into five elements with four quadrature points per element. The grid is predefined before training begins and includes the points where the FENNM outputs are evaluated to construct the total loss function.
\definecolor{darkgreen}{rgb}{0.0, 0.5, 0.0}
\definecolor{tabblue}{rgb}{0.1216, 0.4667, 0.7059}
\newcommand{\bluebullet}{\raisebox{0pt}{\tikz{\filldraw[tabblue] (0,0) circle (1pt)}}}
\newcommand{\greenhollowsquare}{\raisebox{0pt}{\tikz{\draw[darkgreen, line width=0.5pt] (0,0) rectangle (4pt,4pt)}}}
\newcommand{\blackplus}{\raisebox{0pt}{\tikz{\draw[black, line width=0.5pt] 
    (-2pt, 0pt) -- (2pt, 0pt) 
    (0pt, -2pt) -- (0pt, 2pt); 
}}}
\newcommand{\blackx}{\raisebox{0pt}{\tikz{\draw[black, line width=0.5pt] 
    (-2pt, -2pt) -- (2pt, 2pt) 
    (2pt, -2pt) -- (-2pt, 2pt); 
}}}
\newcommand{\redtriangle}{\raisebox{0pt}{\tikz{\filldraw[red] 
    (0pt, 2pt) -- (-2pt, -2pt) -- (2pt, -2pt) -- cycle; 
}}}
\begin{figure}
    \centering
    \includegraphics[width=1.0\linewidth]{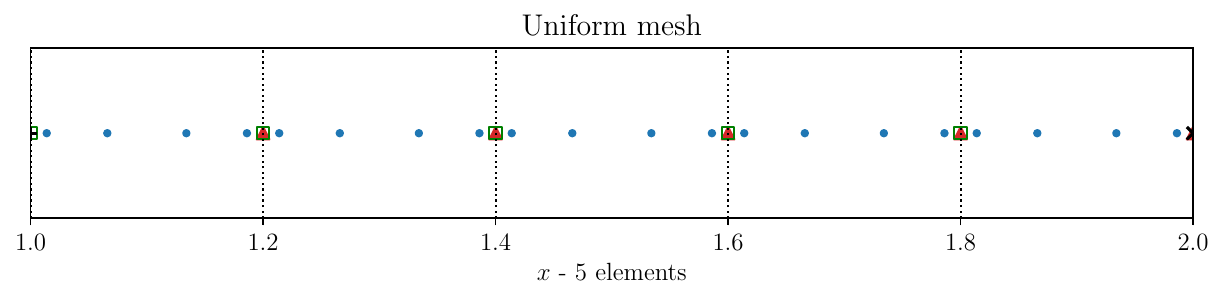}
    \caption{An illustration of a uniform mesh comprising five elements over the domain $x \in [1,2]$. Markers indicate: \protect\bluebullet \ the quadrature points $x_q$ distributed throughout the elements in the global coordinates; \protect\blackplus \ left boundary condition; \protect\blackx \ right boundary condition; \protect\redtriangle \ elements' right boundaries; \protect\greenhollowsquare \ elements' left boundaries.} 
    \label{uniform_mesh}
\end{figure}

The NN output goes through automatic differentiation (AD) to compute the flux values at the elements' boundaries, and the differential operators of the weak-form equation and the forcing term at the quadrature points of the elements in the computational domain \cite{baydin2018automatic}. These outputs are termed convolution signals. Then, the predefined filters pass over their corresponding signals to evaluate the Gauss quadrature sums for each element for each test function in the convolution process. After that, the convolution outputs are grouped to formulate the residual loss tensor for all elements for all test functions. Finally, the tensor is squared element-wise, summed, and averaged over the number of elements and test functions to construct the total residual loss function as in equation (\ref{FENNM_residual_eq}). 
\section{Numerical Experiments and Discussions}\label{discussion}
In the current study, the penalty terms ${\tau_\mathcal{R}}$ and ${\tau_\mathcal{B}}$  in the total loss function in equation (\ref{FEM_eq_PINN}) are nondecreasing variables and are updated simultaneously during the training process using ADAM optimizer \cite{kingma2014adam} and remain constant during optimization using L-BFGS \cite{liu1989limited}. Loss terms with increasing errors are automatically weighted more, forcing the network to minimize them. Hence, the network seeks to find a saddle point where it optimizes its parameters during training using gradient descent to minimize total loss and updates the penalty terms using gradient ascent to maximize their weights \cite{mcclenny2023self, liu2021dual}. 

Table \ref{table1} provides the NN architecture, parameters, and the choice of test functions for case studies from Section \ref{quad_pts} to Section \ref{steep_section}.  Section \ref{quad_pts} explores the residual loss function of FENNM in detail, aiming to establish a link between the necessary quadrature points per element and the components of the residual loss function. The relation between the order of the test function and the convergence rate of FENNM is demonstrated in Section \ref{conv_rate}.

\begin{table}
\caption{FENNM design parameters for case studies from Section 
\label{table1}
\ref{quad_pts} to Subsection \ref{steep_section}.}
\resizebox{\columnwidth}{!}{%
\begin{tabular}{cccccccccc}
\hline
Section & Case study & \begin{tabular}[c]{@{}c@{}}Number of \\neurons/layer\end{tabular} & \begin{tabular}[c]{@{}c@{}}Number \\ of layers\end{tabular} & \begin{tabular}[c]{@{}c@{}}Activation\\  function\end{tabular} & \begin{tabular}[c]{@{}c@{}}Number \\ of elements\end{tabular} & \begin{tabular}[c]{@{}c@{}}Test \\ functions\end{tabular} & \begin{tabular}[c]{@{}c@{}}Quadrature \\ points/element\end{tabular} & \multicolumn{2}{c}{Epochs} \\ \hline
 &  & $\mathcal{N}$ & $\ell$ & $\sigma$ & $n$ & $\phi$ & $Q$ & ADAM & L-BGFS\\ 
 \hline
4.1 & \begin{tabular}[c]{@{}c@{}}Equilibrium\\  equation\end{tabular} & 20 & 2 & tanh & 1 & \begin{tabular}[c]{@{}c@{}}Linear \\ Quadratic\\  Cubic\\  Quartic\end{tabular} & 1-15 & 5000 & 5000 \\
\hline
4.2 & \begin{tabular}[c]{@{}c@{}}Equlibrium\\  equation\end{tabular} & 20 & 2 & tanh & 1-1000 & \begin{tabular}[c]{@{}c@{}}Linear\\ Quadratic\\  Cubic\end{tabular} & \begin{tabular}[c]{@{}c@{}} 3\\4\\5 \end{tabular}& 5000 & 5000 \\
\hline
4.3.1 & Beam & 20 & 3 & tanh & 4 & Cubic Hermite & 5 & 5000 & 10000 \\
\hline
4.3.2 & oscillator & 20 & 4 & sin & 25 & \begin{tabular}[c]{@{}c@{}} Linear \\Quadratic\\ Cubic \end{tabular} & \begin{tabular}[c]{@{}c@{}}3\\4\\5 \end{tabular} & 10000 & 10000 \\
\hline
4.3.3 & Transport & 20 & 4 & tanh & 22/15/11 & Quartic & 10 & 10000 & 10000 \\
\hline
4.3.4 & Poisson & 20 & 4 & sin & 30/14 & Quartic & 10 & 10000 & 10000 \\ \hline
\end{tabular}%
}
\end{table}
\subsection{The Influence of The Test Function Order on The Element Error}
\label{quad_pts}
\begin{figure}
    \centering
    \resizebox{0.8\textwidth}{!}{\includegraphics[]{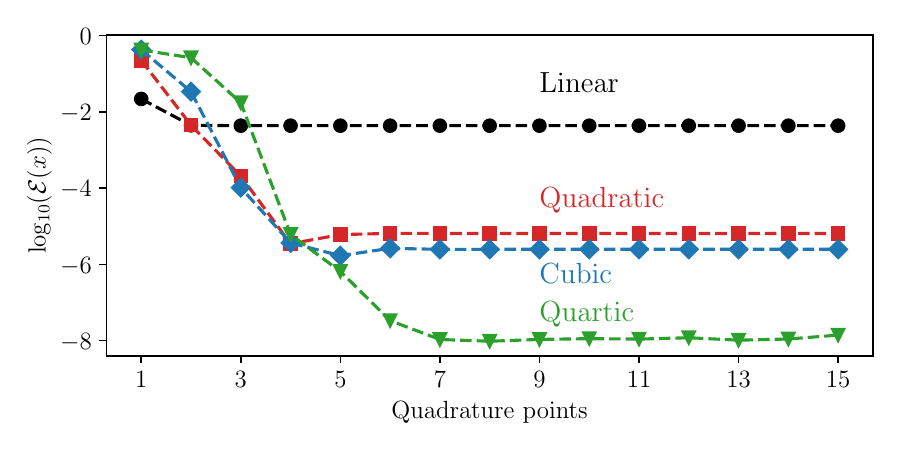}}
    \caption{The average absolute error $\mathcal{E}(x) = 1/N\sum_{i=1}^{N}|U_{NN}(x)-U(x)|$ in FENNM when increasing the number of the quadrature points for different test function orders.}
    \label{quad_pts_study}
\end{figure}

The domain of equation (\ref{FENNM_residual_eq}) is discretized into an element to investigate the relationship between the average absolute error and the number of quadrature points used for a certain order of test functions. Figure \ref{quad_pts_study} shows the average absolute error in FENNM approximation using different orders of Lagrange test functions for a range of quadrature points for a domain consisting of one element. As anticipated, increasing the order of the test function improves the FENNM approximation. However, after a certain number of quadrature points depending on the order of the test function used, the FENNM performance saturates for all test function orders. 

The minimum number $q$ of quadrature points to accurately approximate an integral of a polynomial of order $p$ is defined as $q \geq (p+1)/2$, where $q$ is also the order of the Gauss quadrature rule \cite{burnett1987finite}. To understand the behavior of FENNM in Figure \ref{quad_pts_study}, the order of the Gauss quadrature rule must be identified for the two sums in equation (\ref{FENNM_residual_eq}). The test functions and their derivatives are polynomials of known order. A forcing term of any kind can be used, and it is important to utilize an adequate number of $q$ points to approximate this term accurately. This precision is crucial for the network to minimize the residual loss. Although the network forms a nonlinear solution space to approximate the differential equation operators, the specific trial solution it produces is not accessible. Consequently, the required number of $q$ points for their approximation depends on the particular problem being addressed. Hence, a lower limit of $q$ points can be identified to ensure a good approximation of the residual loss. 
 
 The upper limit is not fixed because it depends on the nature of the forcing term and the complexity of the solution approximated by the FENNM. After identifying the upper limit by trying a higher number of $q$ points, using additional Gauss points is unnecessary since they do not influence the accuracy of FENNM. The number $q$ can be reduced by discretizing the domain into smaller elements, since the complexity of the solution is divided among the elements. This is discussed in Section \ref{conv_rate}. Figure \ref{Error_ratio_all} in appendix \ref{q_p} shows how the FENNM error saturates when using a high number of $q$ points for different test function orders.
\subsection{Rate of Convergence of FENNM}\label{conv_rate}
\begin{figure}
    \centering
    \resizebox{0.8\textwidth}{!}{\includegraphics[]{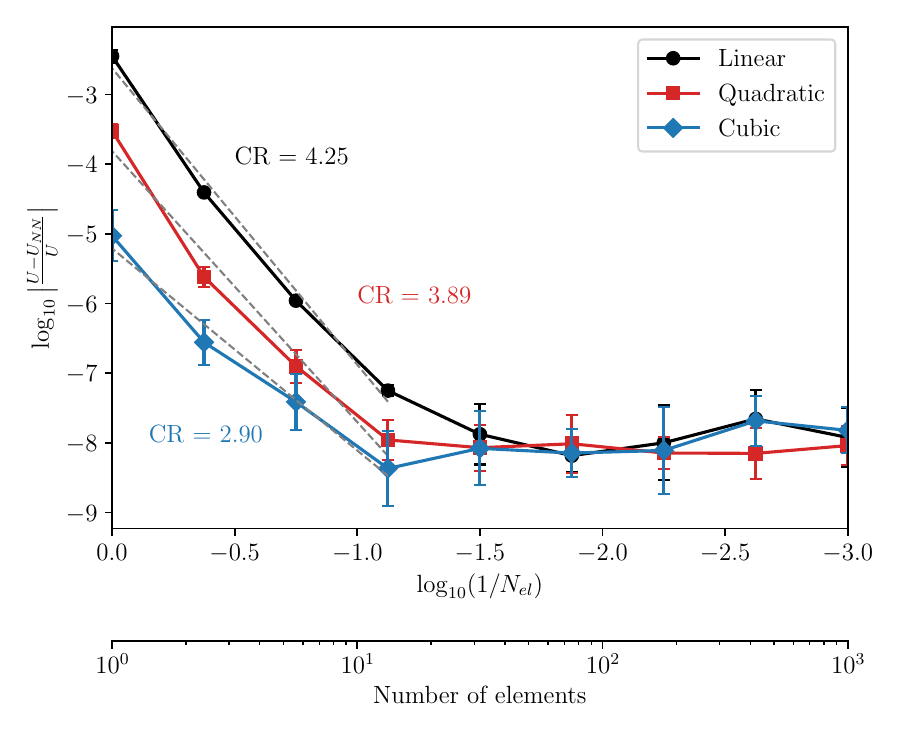}}
    \caption{The rate of convergence of FENNM using linear, quadratic, and cubic test functions at $x=1.5$. The error bars represent 95\% confidence interval, calculated from ten randomly initialized networks for each mesh.}
    \label{conv_rate_plot1}
\end{figure}

The convergence rate (CR) of FENNM was examined through a set of numerical experiments involving equation (\ref{FENNM_residual_eq}), using varying mesh densities and test function orders. Figure \ref{conv_rate_plot1} shows the relative absolute error of FENNM at $x=1.5$ displayed on a log-log scale for linear, quadratic, and cubic test functions, using mesh densities ranging from [1, 1000] elements. 

The CR in Figure \ref{conv_rate_plot1}  is defined as the slope of the curve before it plateaus. An adequate quadrature rule order was employed for each test function based on the study conducted in Section \ref{quad_pts}. The results show that the CR decreases for high-order test functions. A similar trend was observed in a previous study conducted in VPINN to understand the role of the order of the quadrature rule and the test functions \cite{berrone2022variational}. The authors concluded counterintuitively that for a fixed quadrature rule, the best CR is achieved using test functions of the lowest order for smooth solutions \cite{berrone2022variational}. 

A high-order quadrature rule is necessary when employing high-order test functions, as discussed in Section \ref{quad_pts}. Conversely, applying a high-order quadrature rule to low-order test functions indicates that the CR is influenced by the test function order for a given quadrature rule order and network parameters. For coarse meshes, the application of the \textit{ p} refinement has a greater influence on the network accuracy than \textit{h} refinement. However, as the mesh becomes denser, the accuracy of FENNM for different test functions will become similar. FENNM achieved its lowest error in approximating equation (\ref{FENNM_residual_eq}) with linear test functions with an 80-element mesh, while the minimum error for quadratic and cubic test functions was systematically found in 30-element meshes. This error represents a pointwise error (PWE) of the order $\mathcal{O}(-8)$. FENNM reaches this value when the mean square error of the training loss reaches a value below $2.220\times 10^{-16}$, which corresponds to the machine precision of TensorFlow float64, and the network cannot go beyond it. A comparison of the CR of FENNM with that of FEM is shown in \ref{CR_FENNM_vs_FEM}.
\subsection{Numerical Case Studies}\label{numerical}
This section investigates four case studies that evaluate the performance and accuracy of the FENNM. The details of each NN are summarized in Table \ref{table1}. The loss function was initially optimized using the ADAM optimizer \cite{kingma2014adam} for a specified number of iterations to adjust its penalty terms. Subsequently, the L-BFGS optimizer took over the optimization with another specified number of iterations to refine the network parameters \cite{liu1989limited}.

\subsubsection{Cantilever Beam Subjected to An Intermediate Static Force}
\begin{figure}
    \centering
    \resizebox{.8\textwidth}{!}{\includegraphics[]{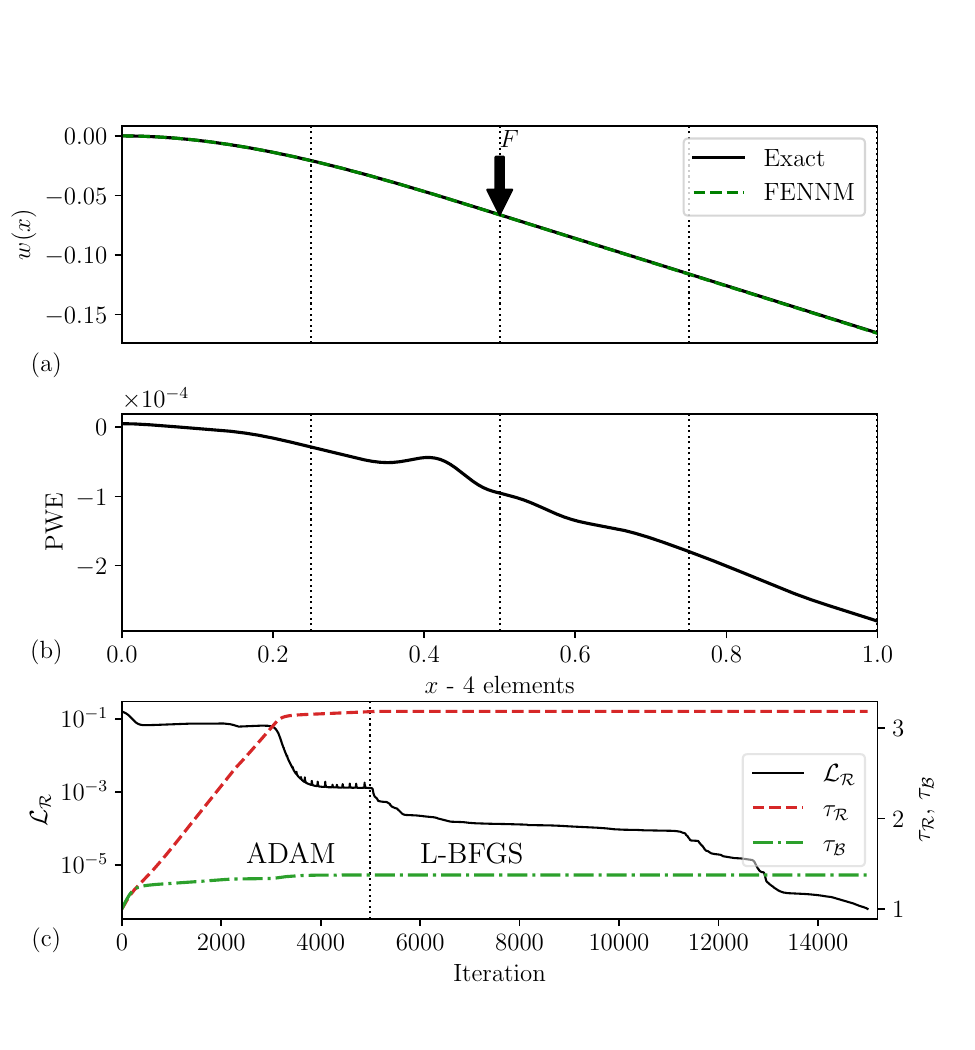}}
    \caption{Solution of a cantilevered beam subjected to an intermediate static force using FENNM: (a) FENNM approximation compared to the exact solution; (b)  PWE over the beam length; (c) training history of the FENNM. The beam is of length $L=1 [m]$, modulus of elasticity $E=69.9\times 10^9 [GPa]$,  second moment of  area of $I=9\times 10^{-9}[m^4]$, and a point load $F=100 [N]$.}
    \label{Beam_4E}
\end{figure}

Consider an Euler-Bernoulli cantilevered beam of length $L$, bending rigidity $EI$, and subjected to a point force $F$ at its midpoint. Its deflection $w(x)$ varies along the position $x$ to obey the differential equation
\begin{equation}\label{beam_eq}
    EI\frac{d^4w(x)}{dx^4} = F\delta(x-\frac{L}{2}),
\end{equation}
with Dirichlet  boundary conditions
\begin{equation}\label{beam_DBC}
    w|_{x=0} = \left. \frac{dw}{dx}\right|_{x=0} = 0 ,
\end{equation}
and Neumann boundary conditions 
\begin{equation}\label{beam_NBC}
    \left. \frac{d^2w}{dx^2}\right|_{x=L/2} = \left. \frac{d^3w}{dx^2}\right|_{x=L} = \left. \frac{d^2w}{dx^2}\right|_{L} = 0.
\end{equation} 
Consider the domain of equation  (\ref{beam_eq})  to be discretized into four elements as shown in Figure \ref{Beam_4E} (a). The residual loss function is
\begin{equation}\label{beam_weakform}
\begin{split}
\mathcal{L}_\mathcal{R} &=\frac{1}{N_{el}K} \sum_{n=1}^{4} \sum_{k=1}^{K^{(n)}} \Bigg( 
    EI \left. \frac{\partial^3 w(x)}{\partial x^3} v_k(\xi) \right|_{x_{n-1}}^{x_n} 
    - EI \frac{1}{J_x} \left. \frac{\partial^2 w(x)}{\partial x^2} 
    \frac{\partial v_k(\xi)}{\partial \xi} \right|_{x_{n-1}}^{x_n} \\
&\quad + EI \sum_{q=1}^{Q} W_q \frac{1}{J_x} 
    \frac{\partial^2 w(x_q^{(n)})}{\partial x^2} 
    \frac{\partial^2 v_k(\xi_q^{(n)})}{\partial \xi^2} 
\Bigg)^2,
\end{split}
\end{equation}
where $v(\xi)$ are the cubic Hermite interpolators in the local coordinates, typical of beam elements \cite{reddy1993introduction}. Breaking down the total loss per element to
\begin{equation}
    \mathcal{L}_\mathcal{R} = \sum_{n=1}^4 \mathcal{L}_\mathcal{R}^{(n)},
\end{equation}
the Neumann boundary conditions in equation (\ref{beam_NBC}) can be imposed into the residual loss function equation (\ref{beam_weakform}) as follows
\begin{subequations}
\begin{align}
\begin{split}
\mathcal{L}_\mathcal{R}^{(2)} &= \sum_{k=1}^{K^{(2)}} \Bigg( 
    EI \left( \left.\frac{-F}{EI} v_k\right|_{\xi = 1}
    - \left.\left.\frac{\partial^3 w}{\partial x^3}\right|_{x_1} v_k\right|_{\xi = -1} \right) - EI \frac{1}{J_x} \left( 0 \cdot \left.\frac{\partial v_k}{\partial \xi} \right|_{\xi =1}
    - \left. \left.\frac{\partial^2 w}{\partial x^2}\right|_{x_1} \frac{\partial v_k}{\partial \xi}\right|_{\xi = -1} \right) \\
&\quad + EI \sum_{q=1}^{Q} W_q \frac{1}{J_x} 
    \frac{\partial^2 w(x_q^{(2)})}{\partial x^2} 
    \frac{\partial^2 v_k(\xi_q^{(2)})}{\partial \xi^2} 
\Bigg)^2,
\end{split} \label{beam_bc1}\\
\begin{split}
\mathcal{L}_\mathcal{R}^{(4)} &= \sum_{k=1}^{K^{(4)}} \Bigg( 
    EI \left( 0 \cdot v_k |_{\xi = 1}
    -\left. \frac{\partial^3 w(x_{3})}{\partial x^3}\right|_{x_3} v_k|_{\xi = -1} \right) - EI \frac{1}{J_x} \left( 0 \cdot \left.\frac{\partial v_k}{\partial \xi} \right|_{\xi =1}
    - \left.\frac{\partial^2 w}{\partial x^2}\right|_{x_3} \left.\frac{\partial v_k}{\partial \xi}\right|_{\xi=-1} \right) \\
&\quad + EI \sum_{q=1}^{Q} W_q \frac{1}{J_x} 
    \frac{\partial^2 w(x_q^{(4)})}{\partial x^2} 
    \frac{\partial^2 v_k(\xi_q^{(4)})}{\partial \xi^2} 
\Bigg)^2.
\end{split} \label{beam_bc2}
\end{align}
\end{subequations}

The intermediate Neumann boundary conditions are imposed at the second element in equation (\ref{beam_bc1}), and the end Neumann boundary conditions are imposed at the fourth element in equation (\ref{beam_bc2}).  Ultimately, the loss function contains the Neumann boundary conditions at their corresponding location, similar to the FEM formulation.  However, unlike the FEM, Dirichlet boundary conditions are defined in the boundary loss function in the FENNM.

Figure \ref{Beam_4E}  shows the solution of the cantilevered beam subjected to an intermediate static force with FENNM as described in equation (\ref{beam_eq}). The deflection $w(x)$ of the beam is compared with the exact solution in Figure \ref{Beam_4E} (a), where the PWE is illustrated in Figure \ref{Beam_4E} (b). The PWE accumulates over the length of the beam because the Neumann boundary conditions are weakly enforced within the residual loss function and remain in order $\mathcal{O}(-4)$.

The penalty terms $\tau_{\mathcal{R}}$ for residual loss and $\tau_{\mathcal{B}}$ for Diriclet boundary conditions in Figure \ref{Beam_4E} (c) increased during the first 5000 iterations with ADAM optimizer. Approximately 3000 iterations later, the network found a saddle point where the penalty values saturated, and simultaneously the residual loss $\mathcal{L}_{\mathcal{R}}$ started to decrease. A direct relation was found between the saturation of the penalty terms and the number of layers $\ell$ used in the NN. When the NN is not deep enough, the penalty terms do not saturate regardless of the number of ADAM iterations used.  Additionally, it was found that the ADAM optimizer struggles with stiff problems and a similar pattern was observed. After that, training was continued with the L-BFGS optimizer for 10,000 iterations to fine-tune the network parameters and converge to the local minimum with the penalty terms remaining constant. 

Employing nonvanishing test functions allows FENNM to integrate intermediate loads into the residual loss function using a single network while maintaining high accuracy across the entire domain. Solving the problem of equation (\ref{beam_eq}) with a vanilla PINN is not straightforward as considering a punctual force would require a piecewise solution with two coupled PINNs. 
\subsubsection{Nonlinear Pendulum}
The FENNM approach can be extended to the temporal dimension in transient problems using the same formulation by treating time as a spatial dimension \cite{kharazmi2021hp}. Consider the nonlinear pendulum equation 
\begin{equation}\label{pendulum_eq}
    \ddot{\theta} + c\dot{\theta} + \frac{g}{L}sin(\theta) = 0,
\end{equation}
with initial conditions
\begin{equation}
    \theta|_{t=0} = \frac{3\pi}{8},\ \dot\theta |_{t=0}= 0,
\end{equation}
where $\theta\ , \dot{\theta}\ ,\text{and } \ddot{\theta}$  are angular displacement, angular velocity, and angular acceleration, respectively. With a damping coefficient $c$, length $L$, and gravitational acceleration $g$. The first-order initial condition can be imposed inside the flux term of the weak-form residual loss function as shown in equation (\ref{pendulum_weak}). In contrast, the zeroth-order initial condition is separately imposed inside the boundary loss function.
\begin{equation}\label{pendulum_weak}
\mathcal{L}_{\mathcal{R}} =  \frac{1}{N_{el}K} \sum_{n=1}^{N_{el}} \sum_{k=1}^{K^{(n)}} 
\left(
\dot{\theta} v_k \big|_{t_n}^{t_{n+1}} 
- \sum_{q=1}^{Q} W_q \dot{\theta} \dot{v}_k 
+ \sum_{q=1}^{Q} J_t W_q \left( c \dot{\theta} + \frac{g}{L} \sin(\theta) \right) v_k
\right)^2,
\end{equation}

\begin{figure}
    \centering
    \resizebox{1.\textwidth}{!}{\includegraphics[]{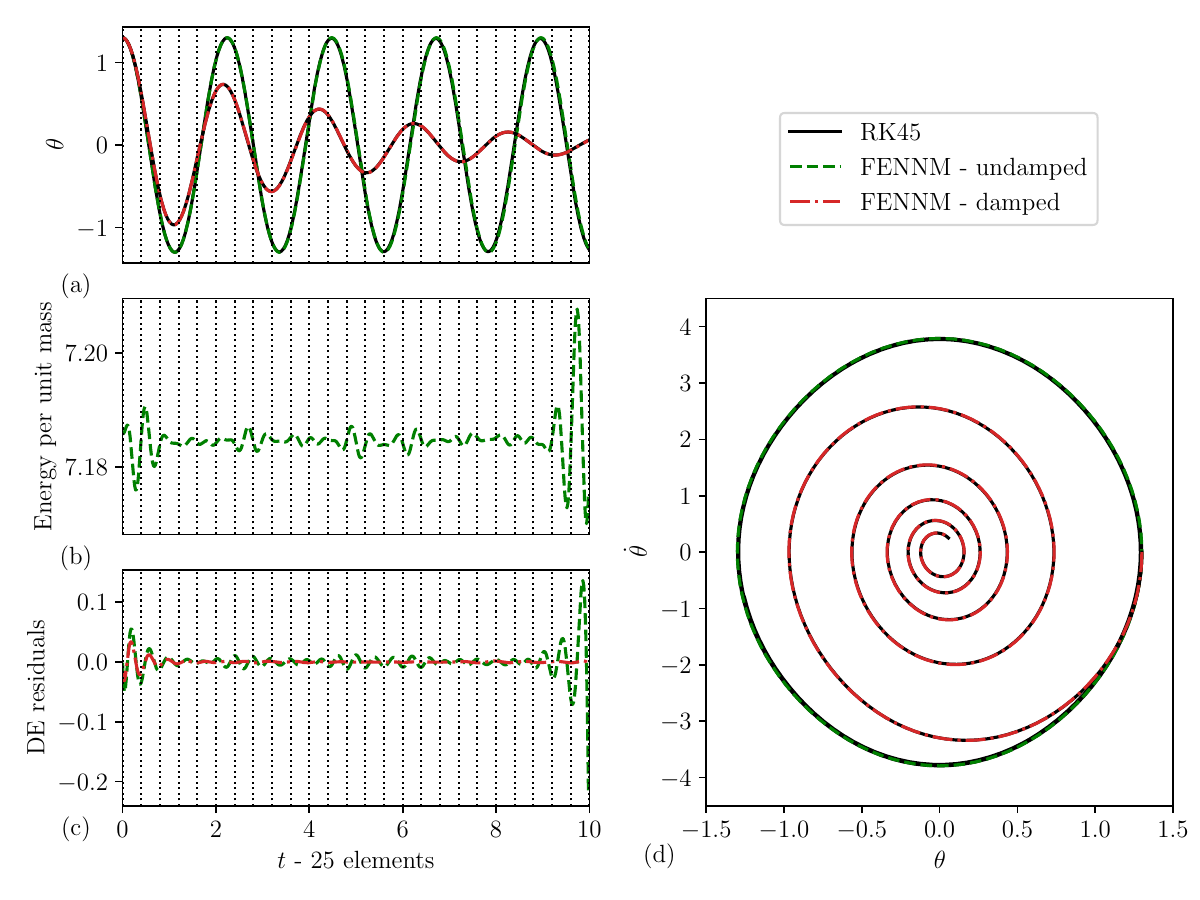}}
    \caption{The solution of damped and undamped nonlinear pendulum using FENNM: (a) FENNM approximation compared to the RK45 method in Python; (b) the total system energy of the undamped pendulum; (c) DE residuals of the FENNM; (d) phase-space plots. Both pendulums have a length $L=1\  [m]$, and a damping coefficient $c=0.5\  [s^{-1}]$.}
    \label{pendulum_plot}
\end{figure}

The solutions of the damped and undamped pendulum in equation (\ref{pendulum_weak}) are shown in Figure \ref{pendulum_plot} computed via FENNM. Figure \ref{pendulum_plot} (a) shows the different oscillations for both pendulums and (d) shows their phase-space plots. 
The FENNM approximation was compared with the Runge-Kutta method of precision of the $4^{th}$ order with steps taken from the $5^{th}$ order (RK45) in Python. Thus, the numerical errors inherent in the RK45 solution prevent an accurate evaluation of the PWE for FENNM. Hence, the total energy of the undamped pendulum is used to assess whether FENNM maintains the conservation of the total energy as presented in Figure \ref{pendulum_plot} (b). It is noticed that the average total energy per unit mass is conserved. However, it oscillates within the elements because of the approximation of the weak-form residuals using the Gaussian quadrature rule. To mitigate such oscillations, one can use more elements or increase the order of the test functions in the weak-form formulation. The residuals of the strong-form equation (\ref{pendulum_eq}) for both pendulums are illustrated in Figure \ref{pendulum_plot} (c) where the DE residuals of the damped pendulum die over time due to the presence of damping in the system. However, the residuals of the undamped pendulum oscillate within the elements with larger magnitudes at both ends of the temporal domain. These oscillations behave similarly to the total energy per unit mass distribution functions depicted in Figure \ref{pendulum_plot} (b) and are a function of the size of the elements. This residual quantity is accessible from the automatic differentiation and can serve as a proxy for the error when performing mesh refinement. 

Segmenting the temporal domain into distinct, nonoverlapping time elements, with each element representing a different time step, and then integrating over the entire temporal domain, allows FENNM to approximate the solution across the computational domain without committing to any specific time integration scheme. 

\subsubsection{The Transport Equation}
The nondimensional transport equation sometimes referred to as Convection-Diffusion equation is given by
\begin{subequations}\;\label{transport_eq}
\begin{align}
\begin{split}
    P_e\frac{du}{d\hat{x}} - \frac{d^2u}{d\hat{x}^2}  &= \hat{f}(\hat{x}), \text{  in  } \Omega, 
\end{split} \label{transport_general_eq} \\
\begin{split}
    u(\hat{x}) &= 0, \text{  on  } \partial \Omega,
\end{split} \label{transport_general_BC}
\end{align}
\end{subequations}
where $\hat{x} = x/L$ is the nondimensional spatial coordinate, and $L$ is the characteristic length of the domain. The Péclet number $P_e = Lv_x/\alpha$ describes the relative importance of convection represented by the velocity of fluid flow $v_x$ to diffusion indicated by the diffusion coefficient $\alpha$, which measures the rate of spreading of the physical quantity \cite{abdollahi2017fluid, larson2013finite}.   The nondimensional forcing term $\hat{f}(x) = f(x)/\alpha$ is defined in the domain $\Omega$ with zero boundary conditions at both ends of the domain $\partial \Omega$. Hence, for small values of $\alpha$, small perturbations of $\hat{f}(x)$ lead to large local values of $du/d\hat{x}$, which introduce thin regions near the boundaries known as layers where the solution undergoes sharp changes \cite{larson2013finite}.  The weak-form loss function takes the form
\begin{equation}\label{transport_weak}
\mathcal{L}_{\mathcal{R}} =  \frac{1}{N_{el}K} \sum_{n=1}^{N_{el}} \sum_{k=1}^{K^{(n)}} 
\left( 
\sum_{q=1}^{Q} W_q \frac{du}{d\hat{x}} \frac{dv_k}{d\xi}
\left.-\frac{du}{d\hat{x}} v_k \right|_{x_n}^{x_{n+1}}
+ \sum_{q=1}^{Q} J_xW_q \left( P_e \frac{du}{d\hat{x}}-f(\hat{x}) \right) v_k
\right)^2,
\end{equation}

\begin{figure}
    \centering
    \resizebox{1.\textwidth}{!}{\includegraphics[]{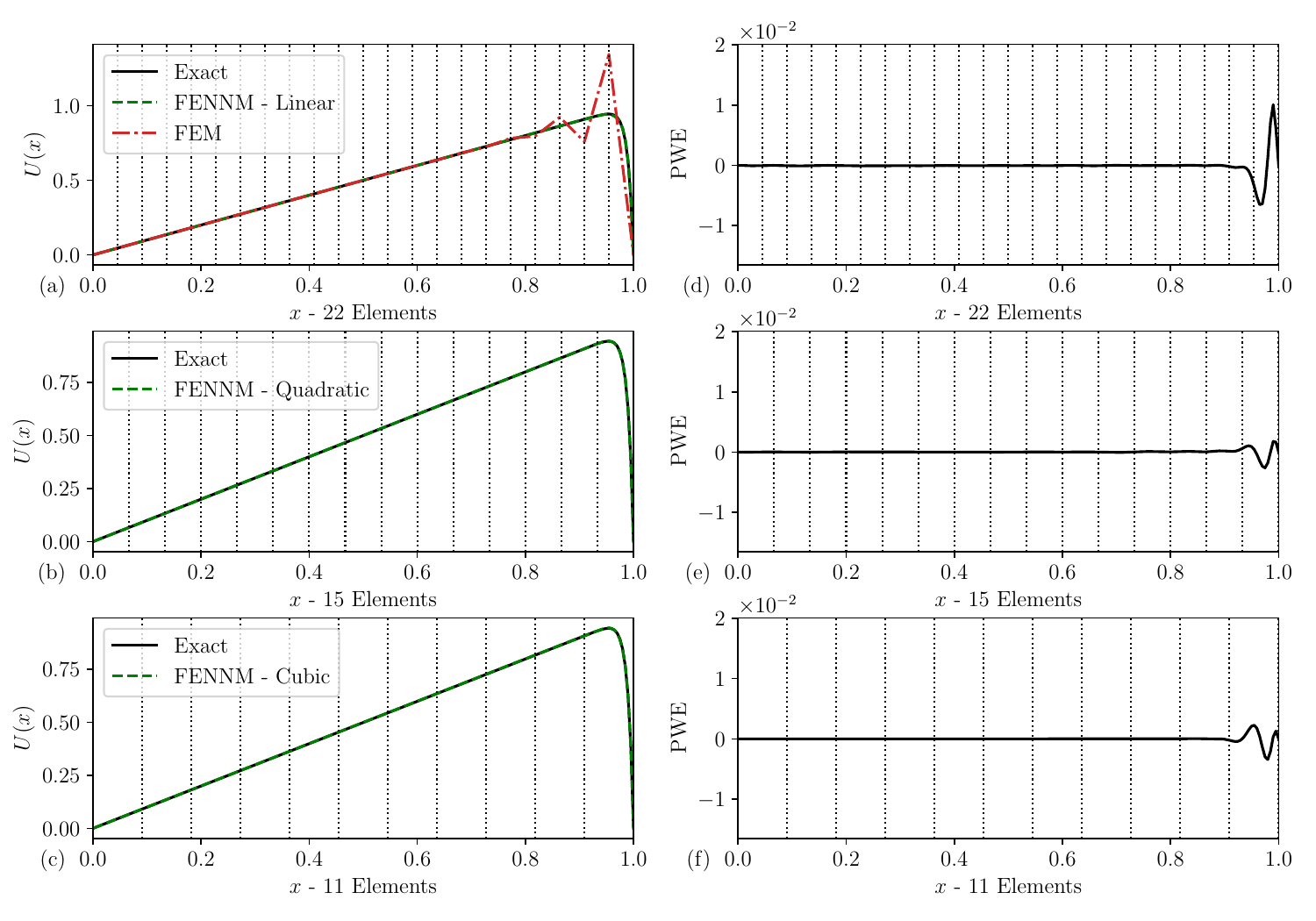}}
    \caption{The solution of transport equation using FENNM: (a) comparison between the solutions of FEM and FENNM compared to the analytical solution using linear test functions in both models; (b and c) FENNM approximations using quadratic and cubic test functions, respectively; (d, e, and f) the PWE of FENNM using linear, quadratic, and cubic test functions, respectively. The domain is of length $L =1$, fluid flow velocity $v_x = 1$,  diffusion coefficient $\alpha = 0.01$, forcing term $f(x) = 1$, and $u(0) = u(L) = 0$.}
    \label{transport}
\end{figure}

Taking into account various orders of test function and mesh sizes, the solution to equation (\ref{transport_weak}) is derived by FENNM as illustrated in Figure \ref{transport}. A comparison between the FEM and FENNM solutions compared to the analytical solution using a mesh of 22 elements using linear test functions is presented in Figure \ref{transport} (a). As expected, the FEM solution introduces oscillations within the computational domain because the element size is larger than the diffusion coefficient \cite{larson2013finite}. However, FENNM captures the sharp change in solution caused by the layer at the boundary with a uniform PWE except for the last element, as illustrated in Figure \ref{transport} (d). In FEM, the solution is based on solving for the nodal values, in which, for coarse meshes and large Péclet number $Pe$, is more influenced by the convective term which shares the information of the solution between every other node \cite{larson2013finite}. Conversely, FENNM relies on reducing the weighted residuals during training, which is based on equating the convective and diffusive terms within each element mitigating the oscillatory issue associated with FEM.

Using higher-order test functions with FENNM enables the reduction of the mesh sizes for the same PWE values. As depicted in Figure \ref{transport} (b, c, e, and f), the PWE consistently maintained uniformity for both quadratic and cubic test functions for 15 and 11 elements, except for the position of the layer. The PWE can be effectively reduced by increasing the element count at the high PWE location, which is discussed in detail in  Subsection \ref{steep_section}.
\subsubsection{One-Dimensional Poisson's Equation with Asymmetric Steep Solution}\label{steep_section}
\begin{figure}
    \centering
    \resizebox{1.\textwidth}{!}{\includegraphics[]{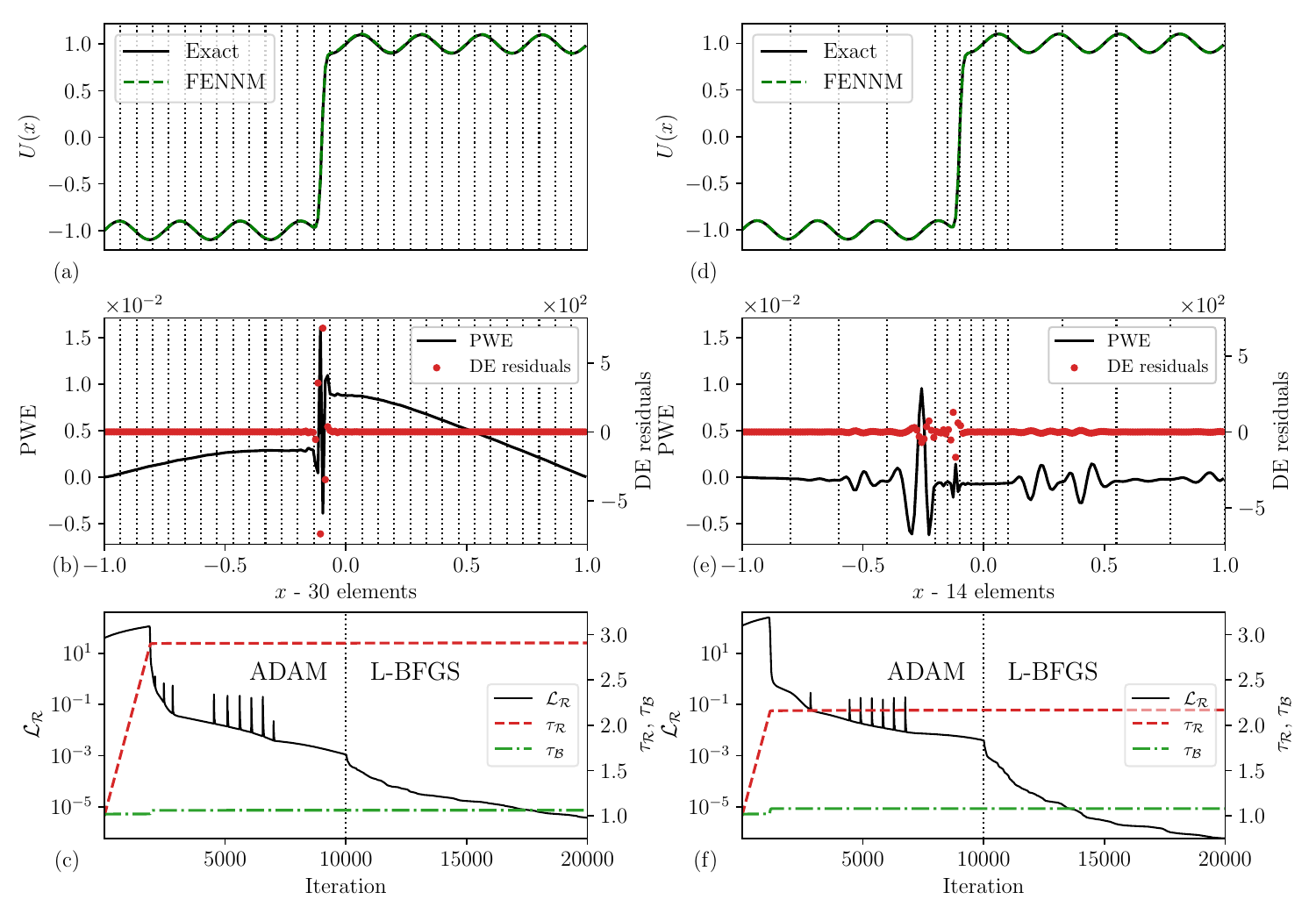}}
    \caption{The solution of one-dimensional Poisson's equation using FENNM: (a and d) the solutions of FENNM compared to the analytical solution using quartic test functions using 30 elements and 14 elements, respectively; (b and e) the PWE and DE residuals of FENNM using 30 elements and 14 elements, respectively; (c and f) the training history of FENNM using 30 elements and 14 elements, respectively.}
    \label{Steep}
\end{figure}

Consider the following Poisson's equation taken from \cite{kharazmi2021hp}
\begin{subequations}\label{steep_eq_total}
\begin{align}
\begin{split}
    - &\frac{d^2U}{dx^2}  = f(x), \text{  in  } \Omega = [-1,1], 
\end{split} \label{steep_general_eq} \\
\begin{split}
    &U(-1) = g, \text{   } U(1) = h, 
\end{split} \label{steep_general_BC}
\end{align}
\end{subequations}
where $g$ and $h$ are constants and $f(x)$ is a forcing term applied over the computational domain. The weak-form loss function takes the general form
\begin{equation}\label{weakformsteep}
    \mathcal{L}_{\mathcal{R}} = \frac{1}{N_{el}K} \sum_{n=1}^{N_{el}} \sum_{k=1}^{K^{(n)}} 
\left ( \sum_{q=1}^Q W_q \frac{dU}{dx}\frac{dv_k}{d\xi} - \left.\frac{dU}{dx}v_k \right|_{x_n}^{x_{n+1}}
- \sum_{q=1}^Q W_q J_x f(x)v_k
\right )^2.
\end{equation}
For asymmetric steep solution, a manufactured solution of equation (\ref{steep_eq_total}) can take the form
\begin{equation}\label{analytical_steep}
    U(x) = 0.1  \text{sin}(8\pi x) + \text{tanh}(80(x + 0.1)),
\end{equation}
where the forcing term is obtained by substituting the manufactured solution (\ref{analytical_steep}) into equation (\ref{steep_eq_total}). Figure \ref{Steep} presents the FENNM solution of equation (\ref{steep_eq_total}) for two different meshes with their corresponding PWE and training histories. The FENNM solution is evaluated against the analytical solution employing Lagrange quartic test functions with a mesh of 30 elements, as depicted in Figure \ref{Steep} (a).
The PWE illustrated in Figure \ref{Steep} (b) shows higher errors in the elements corresponding to the location of the sharp asymmetric change. FENNM is trained on the weak-form weighted residuals, while the DE residuals are not used during training. The residuals of the DE shown in Figure \ref{Steep} (b) correlates well with the PWE and can thus be a good proxy to inform mesh refinement.

To improve the quality of the FENNM approximation at the location of the sharp asymmetric change, local mesh refinement is applied to elements with high residuals and replaced the current mesh with a coarser mesh in regions where residuals are low, as illustrated in Figure \ref{Steep} (d and e).  It is noticed that local mesh refinement reduced the PWE at the sharp change location. However, it introduced a prior noise that can be mitigated with additional local refining. Using less than half of the elements of the uniform grid in Figure \ref{Steep} (a and b), with adaptive mesh refinement, the residuals of the DE remain low in the regions before and after the sharp change and are minimized by a factor of five in the steep asymmetric location, as shown in Figure \ref{Steep} (d and e). This permits the use of automatic adaptive mesh refinement for future applications. 

The training history depicted in Figure \ref{Steep} (c and f) indicates that the use of adaptive mesh refinement influences the overall complexity of the loss function. With the same number of iterations, the training loss achieves a lower value and the penalty terms reached saturation faster.

The performance of FENNM is further tested in approximating the solution of Poisson's equation with a boundary layer solution, as presented in \ref{poisson_BL}, and with a discontinuous forcing term inside the computational domain, as shown in \ref{poisson_disc}.

\section{Conclusion}\label{conclusion}
The finite element neural network method (FENNM)  is developed within Petrov-Galerkin method, where the NN provides the global nonlinear space of solutions, while the test functions belong to the Lagrange test function space and have at least one nonvanishing value at the element boundaries. Consequently, the flux information across the elements is now implemented
inside the weak-form loss function.  FENNM offers several key advantages:
    \begin{itemize}
         \item Convolutions are a fast and parallel way to evaluate the Gauss integration in each element. 
         \item Solving for the weak-form of the differential equations reduces the order of the derivatives in the loss function and the associated error. In addition, it reduces the number of \textit{backpropagation} processes, which speeds up the training process compared to vanilla PINN.
         \item Clear lower and upper limits for the number of quadrature points were established. These limits ensure the approximation of the integral terms within the loss function with the lowest computation cost.
         \item Having nonvanishing values at the elements' boundaries enables imposing natural boundary conditions and intermediate boundary conditions such as point force in the residual loss function with just one network, reducing the number of competing terms in the total loss function. This capability can be extended to include domains with different properties within the loss function for the same inputs rather than adding these properties as additional inputs to the network, which would enhance the optimization and generalization of FENNM.
         \item The strong-form of the residual is not used during training and can be evaluated as a posterior to provide a proxy to guide mesh refinement.
         \item It was shown that, in contrast to FEM, FENNM can be used to discretize and solve in the time dimension, which opens opportunities for approximating solutions for space-time problems in higher dimensions.
         \item FENNM takes a step closer to classical FEM, narrowing the gap between machine learning and traditional numerical methods.
     \end{itemize}
      
Future developments may involve broadening this approach to cover two and three dimensions, integrating time and parameter spaces, handling unstructured meshes, and addressing parametric identification challenges.

\section*{Acknowledgements}
We are grateful for the financial and technical support offered by Hydro-Québec and Maya HTT. We acknowledge the financial support of the Natural Sciences and Engineering Research Council of Canada (NSERC) [funding ref. ALLRP 556353-20], InnovÉÉ and IVADO. We are thankful to L. Berthet, M. Hamedi and B. Blais for the insighful discussions.

\section{The Quadrature Rule Order Influence on The Element Error}\label{q_p}
Figure \ref{Error_ratio_all} shows how the FENNM error saturates after using an upper value of the quadrature points $q$ for the linear, quadratic, cubic, and quartic test functions. As explained in section \ref{quad_pts}, the distribution of the error does not change after this upper limit. Using additional quadrature points per element becomes a computation burden for large meshes with high-order test functions.     
\newcommand{\blackline}{\raisebox{2pt}{\tikz{\draw[-,solid,black,line width = 1pt](0,0) -- (10mm,0);}}}
\newcommand{\greenlinedash}{\raisebox{2pt}{\tikz{\draw[-,dashed,draw=darkgreen,line width = 1pt](0,0) -- (10mm,0);}}}
\newcommand{\blacklinedot}{\raisebox{2pt}{\tikz{\draw[-,dotted,draw=black,line width = 1pt](0,0) -- (10mm,0);}}}
\newcommand{\blackbullet}{\raisebox{0pt}{\tikz{\filldraw[black] (0,0) circle (1pt)}}}
\begin{figure}
    \centering
    \resizebox{1.\textwidth}{!}{\includegraphics[]{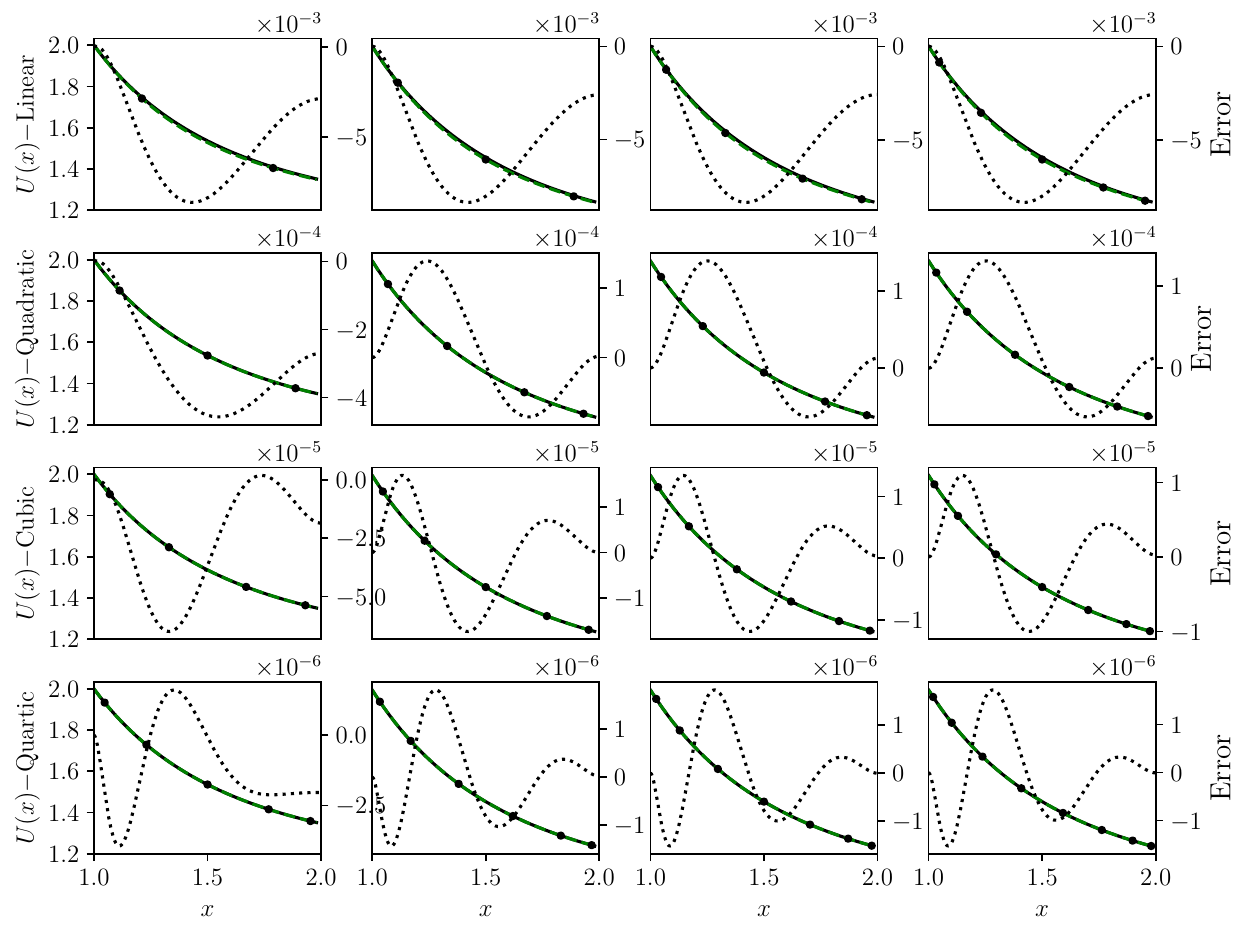}}
    \caption{The error distribution across various quadrature points using linear, quadratic, cubic, and quartic test functions. Markers indicate: \protect\blackline \ analytical solution; \protect\greenlinedash \ FENNM approximation; \protect\blackbullet \ quadrature points; \protect\blacklinedot \ error.}
    \label{Error_ratio_all}
\end{figure}
\section{Rate of Convergence of FENNM Compared to FEM}\label{CR_FENNM_vs_FEM}
\begin{figure}
    \centering
    \resizebox{0.8\textwidth}{!}{\includegraphics[]{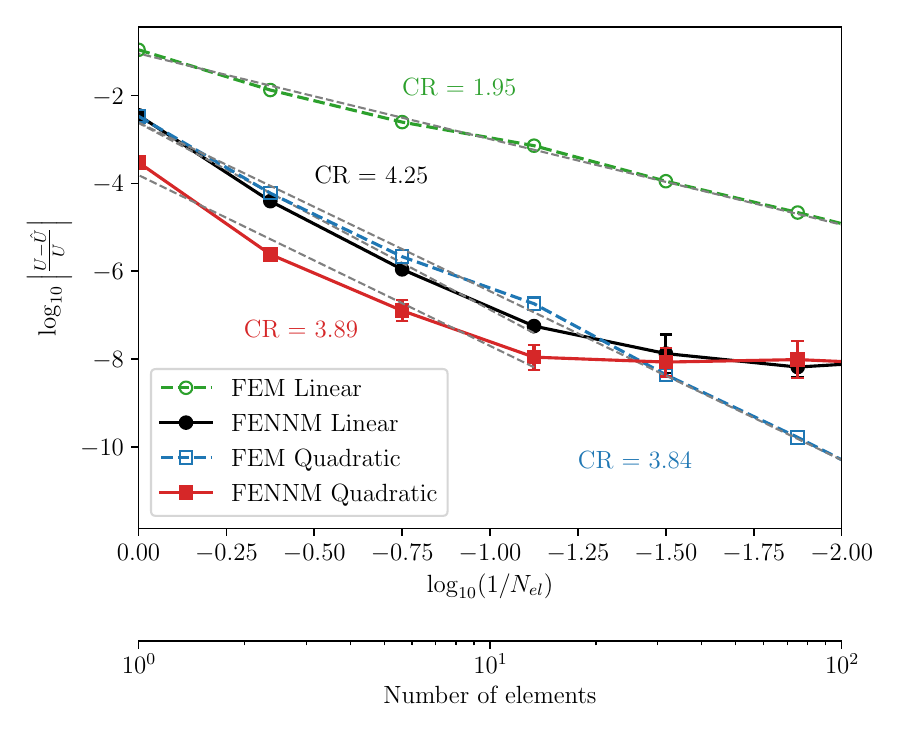}}
    \caption{CR comparison between FENNM and FEM using linear, and quadratic test functions at $x=1.5$. The error bars in FENNM plots represent 95\% confidence interval, calculated from ten randomly initialized networks for each mesh size. In this figure, $\hat{U}$ denotes the trial solution, which applies to both the FEM and FENNM approaches, with its meaning determined by their corresponding labels. The FENNM is a fully connected neural network with $\ell=2$ layers and $\mathcal{N}=20$ neurons per layer with hyperbolic tangent activation functions. }
    \label{conv_rate_plot2}
\end{figure}

Figure \ref{conv_rate_plot2} shows a CR comparison between FEM using the Galerkin framework and FENNM as a Petrov-Galerkin framework for the problem discussed in equation (\ref{FENNM_residual_eq}) using linear and quadratic test functions. As expected, the CR of FEM increases as the order of the test functions increases, especially for dense meshes. 

Unlike FEM where the number of degrees of freedom (DoF) increases as the mesh is refined, the number of DoF in the FENNM model remains the same regardless of the size of the mesh used. The number of DoF of FENNM can be calculated as
\begin{equation}
    \sum_{l=1}^{L}(n_{l-1}.n_l + n_l),
\end{equation}
where $L$ is the number of layers in the NN, $n_{l-1}.n_l$ is the number of weights connecting layer $l-1$ to layer $l$, and $n_l$ is the number of biases in layer $l$ . The number of DoF remains constant in FENNM for all mesh sizes and equals 481 for a network of two hidden layers with 20 neurons per layer, while the number of DoF of the FEM model is a function of the number of elements $N_{el}$ and the order of the test functions $p$ and equals $pN_{el} + 1$. When linear test functions are used, the theoretical CR of the FEM is at a rate of $\mathcal{O}(h^{p+1})$ and is in agreement with the result shown in Figure \ref{conv_rate_plot2}. The accuracy of the FENNM exceeded the accuracy of the FEM for all mesh sizes because the point $x=1.5$ is a regular point \cite{burnett1987finite}. However, the point $x=1.5$ becomes a superconvergent point (zeros of Jacobi polynomials) when the test functions are quadratic. Hence, the theoretical CR of the FEM increases at a rate of $\mathcal{O}(h^{p+2})$, which is in agreement with the result in Figure \ref{conv_rate_plot2} \cite{burnett1987finite}. When the mesh size is 33 elements, both the FENNM quadratic and the FEM quadratic reaches approximately the same error. The double precision of float64 limits the FENNM and since the training loss is squared, the machine cannot go beyond it. However, in the FEM model, the error is not squared and a higher accuracy is achieved at denser meshes. 
\section{FENNM for One-Dimensional Poisson's Equation}
The FENNM performance is further evaluated through two experiments on Poisson's equation: one with a boundary layer solution briefly explained in \ref{poisson_BL}, and the other with a discontinuous forcing term inside the computational domain briefly explained in \ref{poisson_disc}. 
\label{appendix_c}
\subsection{Poisson's Equation with Boundary Layer Solution}\label{poisson_BL}
Consider the Poisson's equation (\ref{steep_eq_total}) with a boundary layer solution taken from \cite{kharazmi2021hp}
\begin{equation}
    U(x) = 0.1 \text{sin}(5\pi x) + e^{100x-99}.
\end{equation}

\begin{figure}
    \centering
    \resizebox{1.\textwidth}{!}{\includegraphics[]{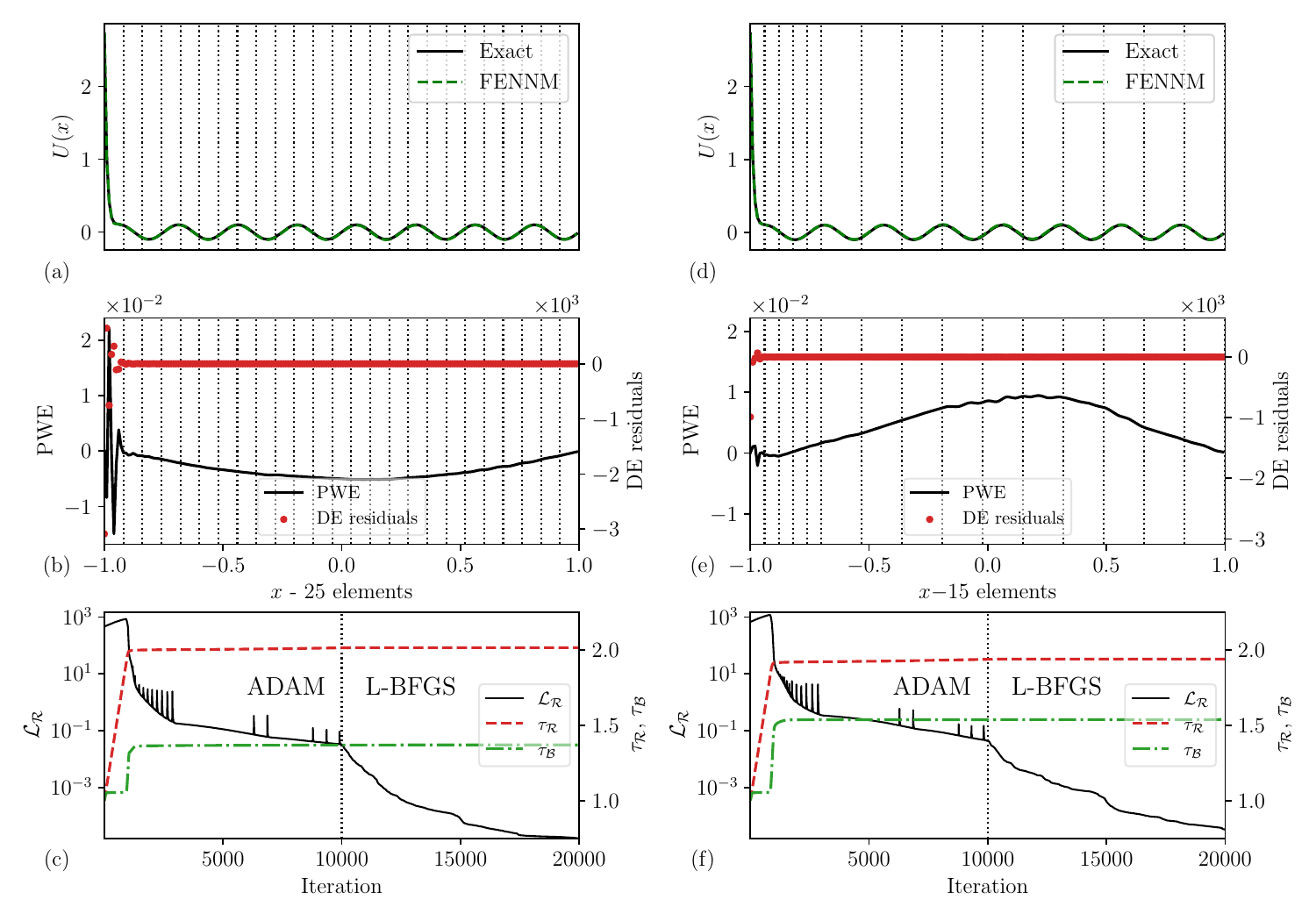}}
    \caption{The solution of one-dimensional Poisson's equation using FENNM: (a and d) the solutions of FENNM compared to the analytical solution using quartic test functions using 25 elements and 15 elements, respectively; (b and e) PWE and DE residuals of FENNM using 25 elements and 15 elements, respectively; (c and f) the training history of FENNM using 25 elements and 15 elements, respectively. Ten quadrature points per element were used for each grid. The FENNM is a fully connected neural network with $\ell=4$ layers and $\mathcal{N}=20$ neurons per layer with sine activation function.}
    \label{poisson_BL_plt}
\end{figure}

Comparing the results in Figure \ref{poisson_BL_plt} (a, b, d, and e), a similar trend is found to that in Section \ref{steep_section}. Adaptive mesh refinement reduced the PWE and the DE residuals even when fewer elements were used. However, although local mesh refinement minimized the large oscillation in the PWE at the location of the boundary layer in Figure \ref{poisson_BL_plt} (e), the PWE increased slightly within the remaining domain. Decreasing this error can be achieved by incorporating more elements throughout the rest of the domain. 
The training history in Figure \ref{poisson_BL_plt} (c and f) indicates similarly that additional elements are required to decrease the training loss when adaptive mesh refinement is applied to the computational domain.

\subsection{Poisson's Equation with Discontinuous Force}\label{poisson_disc}
Consider the following Poisson's equation (\ref{steep_eq_total}) with a discontinuous forcing term given by
\begin{equation}
f(x) =
    \begin{cases} 
  -10 & \text{if } -1 < x < 0, \\
  +10 & \text{if } \ \ \ \ 0 \leq x < 1.
\end{cases}
\end{equation}

\begin{figure}
    \centering
    \resizebox{0.8\textwidth}{!}{\includegraphics[]{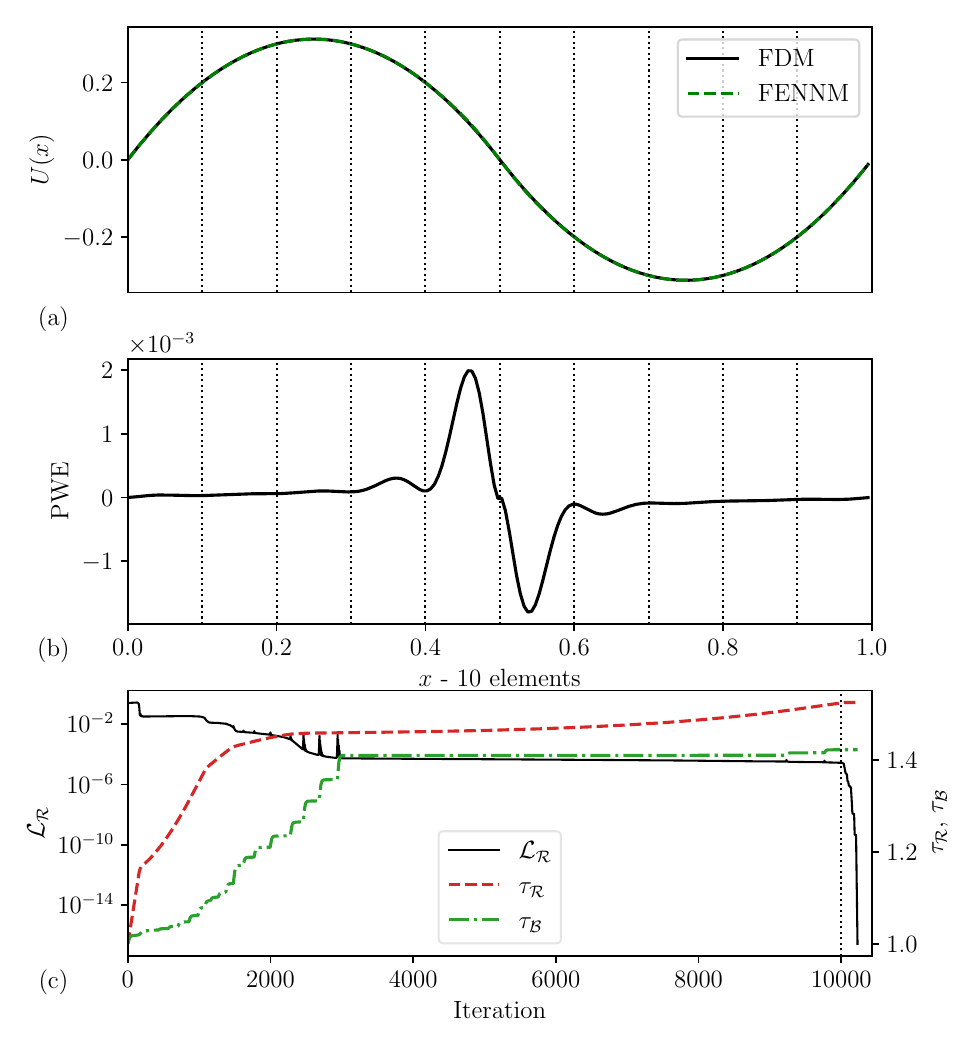}}
    \caption{The solution of one-dimensional Poisson's equation using FENNM: (a) the FENNM approximation compared to FDM, (b) PWE over the computational domain, and (c) the training history of the FENNM. The FENNM is a fully connected neural network with $\ell=4$ layers and $\mathcal{N}=20$ neurons per layer with sin activation function.}
    \label{DiscPoisson_plt}
\end{figure}

This experiment aims to evaluate the capability of FENNM to capture the solution as the forcing term varies throughout the computational domain with a discontinuity. The approximation of FENNM using three quadrature points per element is compared to the finite difference method (FDM) in Figure \ref{DiscPoisson_plt} (a).
In addition, the PWE in Figure \ref{DiscPoisson_plt} (b) oscillates only at the two elements where the forcing term changes its value. This can be mitigated by applying adaptive mesh refinement in that region.  The ADAM optimizer in Figure \ref{DiscPoisson_plt} (c) achieved a low training loss early. However, it saturated for the remaining iterations. Meanwhile, the L-BFGS optimizer minimized the training loss to machine precision with a few iterations afterward.









\renewcommand\bibname{REFERENCES}
\bibliography{References}
\bibliographystyle{IEEEtran}

\end{document}